\documentclass[conference,10pt]{IEEEtran}
\IEEEoverridecommandlockouts
\usepackage{cite}
\usepackage{amsmath,amssymb,amsfonts}
\usepackage{algorithmic}
\usepackage{graphicx}
\usepackage{textcomp}
\usepackage{xcolor}

\usepackage{url} 
\usepackage{optidef}
\usepackage{nicefrac}
\usepackage{mathtools}
\usepackage[caption=false,font=footnotesize]{subfig}
\DeclareMathOperator*{\argmax}{argmax}
\DeclareMathOperator*{\argmin}{argmin}

\begin{document}

\title{A Deep Learning Approach for User-Centric Clustering in Cell-Free Massive MIMO Systems\\
\thanks{This paper was partly supported by the European Union under the Italian National Recovery and Resilience Plan (NRRP) of NextGenerationEU, partnership on “Telecommunications of the Future” (PE00000001 - program “RESTART”, Structural Project 6GWINET, Cascade Call Project SPARKS). 
G. Di Gennaro was supported by the Italian Ministry for University and Research (MUR) - PON Ricerca e Innovazione 2014–2020 (D.M. 1062/2021). 
S. Buzzi was also supported by Horizon Europe project CENTRIC (Grant No. 101096379). 
F.A.N. Palmieri was also supported by POR CAMPANIA FESR 2014/2020, A-MOBILITY: Technologies for Autonomous Vehicles.}
}

\author{\IEEEauthorblockN{Giovanni Di Gennaro}
\IEEEauthorblockA{\textit{Dipartimento di Ingegneria}\\
\textit{Università degli Studi della Campania}\\
Aversa (CE), 81031, Italy\\
giovanni.digennaro@unicampania.it}
\and
\IEEEauthorblockN{Amedeo Buonanno}
\IEEEauthorblockA{\textit{Department of Energy Technologies}\\
\textit{and Renewable Sources, ENEA}\\
Portici (NA), 80055, Italy\\
amedeo.buonanno@enea.it}
\and
\IEEEauthorblockN{Gianmarco Romano}
\IEEEauthorblockA{\textit{Dipartimento di Ingegneria}\\
\textit{Università degli Studi della Campania}\\
Aversa (CE), 81031, Italy\\
gianmarco.romano@unicampania.it}
\and
\IEEEauthorblockN{Stefano Buzzi}
\IEEEauthorblockA{
\textit{University of Cassino and Southern Lazio, I-03043 Cassino, Italy}\\
and \textit{Politecnico di Milano, I-20133 Milano, Italy}\\
buzzi@unicas.it}
\and
\IEEEauthorblockN{Francesco A.N. Palmieri}
\IEEEauthorblockA{\textit{Dipartimento di Ingegneria}\\
\textit{Università degli Studi della Campania}\\
Aversa (CE), 81031, Italy\\
francesco.palmieri@unicampania.it}
}

\maketitle

\begin{abstract}
Contrary to conventional massive MIMO cellular configurations plagued by inter-cell interference, cell-free massive MIMO systems distribute network resources across the coverage area, enabling users to connect with multiple access points (APs) and boosting both system capacity and fairness across user.
In such systems, one critical functionality is the association between APs and users: determining the optimal association is indeed a combinatorial problem of prohibitive complexity. 
In this paper, a solution based on deep learning is thus proposed to solve the user clustering problem aimed at maximizing the sum spectral efficiency while controlling the number of active connections.
The proposed solution can scale effectively with the number of users, leveraging long short-term memory cells to operate without the need for retraining.
Numerical results show the effectiveness of the proposed solution, even in the presence of imperfect channel state information due to pilot contamination. 
\end{abstract}


\section{Introduction}
The issue of inter-cell interference, which often results in subpar performance at the edges of cells in conventional cellular networks, is effectively addressed by Cell-free massive MIMO (CF-mMIMO) \cite{Demir2021}.
This is achieved by removing cell boundaries and utilizing a large quantity of distributed \emph{access points} (APs) that work together to serve a reduced number of \emph{user equipments} (UEs) using the same time-frequency resources. 
The APs are connected via fronthaul to a centralized processing unit (CPU) responsible for coordination and data processing.
However, this setup does not scale efficiently with an increasing number of UEs in terms of required fronthaul capacity and computational complexity \cite{Mendoza2023}. 

To address this issue, a proposal has been made to limit the number of APs that can serve each UE, leading to a user-centric approach to CF-mMIMO \cite{Ammar2022}, \cite{Buzzi2020}. 
In this context, a user-centric cluster represents a subset of APs that are capable of serving a defined UE differently by classic CF-mMIMO where each user can potentially be served by all APs. 
Typically, clusters for each user are formed using specific heuristics that consider various factors, including channel strength, distance, etc \cite{Buzzi2020, Ammar2019, Bjornson2020}.
Optimal AP-UE association is indeed a combinatorial problem whose complexity becomes soon unmanageable even for small size networks.

This paper proposes a \emph{Deep Learning} (DL) approach to perform AP-UE association with the objective of maximizing the network sum spectral efficiency while keeping under control the number of active AP-UE connections. Interestingly, the proposed neural network architecture, being based on \emph{Long Short-Term Memory} (LSTM) network, is scalable with respect to the number of UEs, i.e., it can work with any number of active UEs without the need of being retrained. Numerical results confirm the effectiveness of the proposed approach.

The paper is organized as follows. Next section contains the system model, while in Section III the considered optimization problem is formulated and solved though the proposed deep learning approach. Numerical results are reported in Section IV, while, finally, concluding remarks are given in Section V. 

\section{System model}
We consider a downlink scenario consisting of $K$ single-antenna UEs and $L$ APs geographically located in the same area.
Each AP is assumed to be equipped with $N$ antennas and operates under a massive MIMO regime (i.e., $L \times N \gg K$).

Communication between APs and UEs occurs via \emph{Time-Division Duplex} (TDD) protocol, where each coherence block is used for both uplink and downlink operations.
Specifically, within the coherence block of $\tau_c$ symbols, $\tau_p$ symbols are allocated for uplink pilots, $\tau_u$ symbols for uplink data, and $\tau_d$ symbols for downlink data; ensuring that $\tau_c = \tau_p + \tau_u + \tau_d$.
For the channel model, we adopt \emph{uncorrelated Rayleigh fading} to represent non-line-of-sight (NLoS) communication, with the channel between AP $\ell \in \mathcal{L}=\{1,\dots,L\}$ and UE $k \in \mathcal{K}=\{1,\dots,K\}$ given by $\mathbf{h}_{k\ell} = \sqrt{\beta_{k\ell}} \tilde{\mathbf{h}}_{k\ell} \in \mathbb{C}^N$,
where the variance $\beta_{k\ell}$ represents the large-scale fading (which incorporates shadow fading and geometric path loss), while $\tilde{\mathbf{h}}_{k\ell} \sim \mathcal{N}_\mathbb{C}(\mathbf{0}_N, \mathbb{I}_N)$ accounts for small-scale fading \cite{Demir2021}.

\subsection{Communication process}
To join the network, the $k$-th UE will first identify its \emph{master AP} opting for the one with the strongest channel
\begin{equation}
    m_k = \argmax_{\ell \in \mathcal{L}} \beta_{k\ell}
\end{equation}
where the channel gain measurements are assumed to be obtainable from the synchronization signals periodically transmitted by the APs.
Using these signals, the UE $k$ will then contact its master AP (e.g., via a standard random access procedure) to be assigned a pilot and initiate communication.
The network assumes $\tau_p$ mutually orthogonal pilot sequences $\phi_1, \dots, \phi_{\tau_p}$, where $\|\phi_t\|^2 = \tau_p \quad \forall t \in \mathcal{T}=\{1,\dots,\tau_p\}$.
For the sake of simplicity, we assume that the master AP locally selects pilots for associated UEs.
As specified in \cite{Bjornson2020}, this simply translates to minimizing the interference of the pilots within that AP, by assigning the $k$-th UE the pilot with index $t_k$ defined as
\begin{equation}
	t_k = \argmin_{t \in \mathcal{T}} \sum_{i \in \mathcal{P}_t} \beta_{i m_k}
\end{equation}
where $\mathcal{P}_t \subset \mathcal{K}$ is the set of UEs sharing the $t$-th pilot.

After the pilot assignment, each master AP will communicate its information to the network via the CPU, thereby informing all APs about the newly added UE's details.
Typically, at this stage (centrally or distributedly), each AP evaluates the option of engaging with or abstaining from communicating with the specific UE.
This process determines the set $\mathcal{L}_k \subset \mathcal{L}$ of APs responsible for communicating with the UE $k$, which constitutes the objective of this paper.

\subsubsection{Channel estimation}
Since the TDD protocol is assumed to be synchronized across APs, the transmission of all pilots from UEs connected to the network will consistently occur within the $\tau_p$ samples of the subsequent coherence block.
To execute a coherent transmission, the generic AP $\ell \in \mathcal{L}_k$ should be capable of determining at least an estimate of the channel vector $\mathbf{h}_{k\ell}$ by assessing the received signal $Y_\ell^\mathrm{pilot} \in \mathbb{C}^{N \times \tau_p}$ during that phase, expressed as
\begin{equation}
	Y_\ell^\mathrm{pilot} = \sum_{i=1}^K \sqrt{\eta_i} \mathbf{h}_{i\ell} \phi_{t_i}^\top + \mathbf{N}_\ell \qquad
\end{equation}
where $\mathbf{N}_\ell \in \mathbb{C}^{N \times \tau_p}$ is the noise at the receiver, with i.i.d. elements distributed according to $\mathcal{N}_\mathbb{C}(0, \sigma_{\mathrm{ul}}^2)$, and $\eta_i \ge 0$ is the uplink power of the $i$-th UE.
The channel estimate can be obtained by first removing the interference caused by the orthogonal pilots, i.e., by multiplying it with the normalized conjugate of the corresponding pilot $\phi_{t_k}$, resulting in 
\begin{equation}
    \mathbf{y}_{t_k \ell}^\mathrm{pilot}=
\frac{1}{\sqrt{\tau_p}} Y_\ell^\mathrm{pilot} \phi^{*}_{t_k}
\end{equation}
Subsequent MMSE estimation yields
\begin{equation}
    \mathbf{\hat{h}}_{k\ell} = \frac{\gamma_{k\ell}}{\sqrt{\tau_p \eta_k} \beta_{k\ell}} \mathbf{y}_{t_k \ell}^\mathrm{pilot}
\end{equation}
where, to streamline the upcoming equations, we have defined
\begin{equation}
    \gamma_{k\ell} = \frac{\tau_p \eta_k \beta_{k\ell}^2}{\displaystyle\tau_p\!\!\sum_{i \in \mathcal{P}_{t_k}}\!\!\eta_i \beta_{i\ell} + \sigma_{\mathrm{ul}}^2}
\end{equation}

\subsubsection{Downlink operation}
After the channel estimation phase, downlink transmission takes place. 
The downlink signal received from the UE $k$ can be expressed by
\begin{equation}
    y_k^\mathrm{dl} = \underbrace{\vphantom{\sum_{\setminus}}s_k \sum_{\ell \in \mathcal{L}_k} \mathbf{h}_{k\ell}^\mathrm{H} \mathbf{w}_{k\ell}}_\text{Desired} + 
    \!\!\underbrace{\sum_{i \in \mathcal{K} \setminus \{k\}}\!\!\!\!s_i \sum_{\ell \in \mathcal{L}_i} \mathbf{h}_{k\ell}^\mathrm{H} \mathbf{w}_{i\ell}}_\text{Interference} +
    \underbrace{\vphantom{\sum_{\setminus}}n_k}_\text{Noise}
\end{equation}
where $n_k \sim \mathcal{N}_\mathbb{C}(0, \sigma_\mathrm{dl}^2)$ represents the noise at the receiver, $s_i \in \mathbb{C}$ is the unity power downlink data signal for the generic $i$-th UE ($\mathbb{E}\{|s_i|^2\} = 1$), and $\mathbf{w}_{i\ell} \in \mathbb{C}^N$ is the effective precoding vector.
A feasible option for straightforward distributed implementation is the \emph{Maximum Ratio} (MR) precoding, which for each AP $\ell \in \mathcal{L}_k$ is obtained simply by setting $\mathbf{w}_{k\ell} = \nicefrac{\sqrt{\rho_{k\ell}} \mathbf{\hat{h}}_{k\ell}}{\sqrt{\mathbb{E}\{\|\mathbf{\hat{h}}_{k\ell}\|^2\}}}$, where $\rho_{k\ell} \ge 0$ is the transmit power assigned by AP $\ell$ to UE $k$.
Using a distributed approach to enhance scalability, it is well known that a lower bound to  the downlink \emph{Spectral Efficiency} (SE) for UE $k$ is
\begin{equation}
	\mathrm{SE}_k = \frac{\tau_d}{\tau_c} \log_2 \left(1 + \mathrm{SINR}_k\right) \qquad \text{bit/s/Hz}
\end{equation}
where the pre-log factor $\tau_d / \tau_c$ denotes the portion of each coherence block used for data transmission.
Opting for MR precoding and MMSE channel estimation, the effective \emph{Signal-to-Interference-plus-Noise Ratio} (SINR) is then given by
\begin{equation*}
    \mathrm{SINR}_k = \frac{\displaystyle N \left(\sum_{\ell \in \mathcal{L}_k} \sqrt{\rho_{k\ell} \gamma_{k\ell}} \right)^{\!\!2}}{\displaystyle \sum_{i \in \mathcal{K}} \sum_{\ell \in \mathcal{L}_i} \rho_{i\ell}\beta_{k\ell} + N\!\!\!\!\!\sum_{i \in \mathcal{P}_k \setminus\{k\}}\!\!\!\left(\sum_{\ell \in \mathcal{L}_i} \sqrt{\rho_{i\ell} \gamma_{k\ell}} \right)^{\!\!2} + \sigma_\mathrm{dl}^2}.
\end{equation*}

\section{Problem formulation and DL approach}
Based on the above assumption for calculating the SE, we can compactly formulate the constrained non-convex optimization problem we aim to address as follows
\begin{maxi!}
    {\scriptstyle\mathcal{L}_1, \dots, \mathcal{L}_K}{\sum_{k \in \mathcal{K}} (\mathrm{SE}_k - \lambda |\mathcal{L}_k|) \label{eq:objectiveProblem}}{\label{eq:Problem}}{} \addConstraint{\mathcal{L}_k}{\ne\emptyset}{\forall k \in \mathcal{K} \label{eq:C1Problem}}
    \addConstraint{\sum_{i \in \mathcal{K}_\ell} \rho_{i\ell}}{\leq \rho_\mathrm{max} \qquad}{\forall \ell \in \mathcal{L} \label{eq:C2Problem}}
\end{maxi!}
where $\mathcal{K}_\ell \subset \mathcal{K}$ specifies the set of UEs served by the AP $\ell$, and the parameter $\lambda$ is a pre-defined positive weighting factor (i.e., not learned).
The objective function in \eqref{eq:objectiveProblem} aims to maximize the sum of the SE for all UEs while minimizing the number of active connections, striking a balance between optimizing performance and limiting the number of active~connections.
This problem is known to be NP-hard, since each change in connections alters the SE based on varying channels.
The two constraints presented focus on ensuring a minimum acceptable quality of service and adhering to a physical power limitation.
Constraint \eqref{eq:C1Problem} implies that each UE must be connected to at least one AP, to ensure no UE remains disconnected, and constraint \eqref{eq:C2Problem} guarantees that the total power allocated by each AP does not exceed the maximum transmission power $\rho_\mathrm{max}$, which we assume to be the same for all APs.

\begin{figure}[!t]
    \centering
    \includegraphics[width=0.8\linewidth]{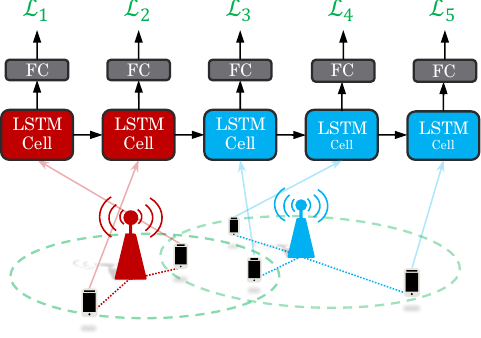}
    \caption{Deep learning framework using UE ordering by master APs.}
    \label{fig:DL}
\end{figure}

In solving the optimization problem \eqref{eq:objectiveProblem}, we aim to design a system that can adapt to variations in the UEs number $K$ without requiring retraining and/or a change in the network structure. 
To achieve this flexibility, we used an LSTM network, which enables the creation of a recursive chain of any desired length.
However, to function effectively, LSTM relies on a strong order of inputs \cite{DiGennaro2021}. 
This requirement does not pose a limitation in real systems where the AP positions are fixed and known to the network operator, enabling simple scheduling among them
(as in \cite{DiGennaro2022}).
This choice proved to be crucial, as learning failed without proper scheduling.

Once prioritized, each AP is associated only with the UEs for which it is the master, creating a first level of hierarchy that places individual UEs within the subchains related to their master APs.
In Fig.~\ref{fig:DL}, this hierarchical level is illustrated by the two distinct colors corresponding to the two APs, sequentially scheduled as first red and then blue.
Each AP will then manage the UEs under its control, sorting them based on channel gain to establish a secondary hierarchical order with a quasi-radial arrangement.
Depending on the number of users to handle, each AP will therefore create an LSTM subchain of an appropriate length, with each cell dedicated to managing a single UE.
For clarity, Fig.~\ref{fig:DL} displays two subchains with lengths of $2$ and $3$, respectively.
It's worth noting that, as the LSTM is a recurrent network, all its cells are identical. 
Consequently, after learning, each cell will consistently respond in the same manner to inputs, making its behavior effectively independent of the number of users.

The inputs to each cell come from information about the individual UE, presumed to be known to the AP through central sensing or the UE's previous uplink transmission.
Specifically, assuming that the $k$-th UE is assigned to the $j$-th cell by its AP master, the input vector to that cell will be
\begin{equation}
    \pmb{\xi}_j = [\beta_{k1}, \beta_{k2}, \dots, \beta_{kL}, x_k, y_k]^\top
\end{equation}
where $x_k$ and $y_k$ are the positions of the UE.
This input is processed within the gates of the LSTM cell, which are defined by the following set of equations
\begin{align*}
    \mathbf{f}_j &= \sigma(\mathbf{W}_f\,\pmb{\xi}_j + \mathbf{U}_f\,\pmb{\upsilon}_{j-1} + \mathbf{b}_f) \\
    \mathbf{i}_j &= \sigma(\mathbf{W}_i\,\pmb{\xi}_j + \mathbf{U}_i\,\pmb{\upsilon}_{j-1} + \mathbf{b}_i) \\
    \mathbf{o}_j &= \sigma(\mathbf{W}_o\,\pmb{\xi}_j + \mathbf{U}_o\,\pmb{\upsilon}_{j-1} + \mathbf{b}_o) \\
    \mathbf{c}_j &= \tanh(\mathbf{W}_c\,\pmb{\xi}_j + \mathbf{U}_c\,\pmb{\upsilon}_{j-1} + \mathbf{b}_c)
\end{align*}
where all the matrices $\mathbf{W} \in \mathbb{R}^{q \times (L + 2)}$ and $\mathbf{U} \in \mathbb{R}^{q \times q}$, along with all the bias vectors $\mathbf{b} \in \mathbb{R}^q$, represent the learned weights from the training phase, while the \emph{output vector} $\pmb{\upsilon}_{j-1} \in \mathbb{R}^q$ is directly influenced by the preceding cell.
Note that the only parameter to be defined in the simulation phase is the dimension $q$ of the LSTM's hidden state.
The output vector $\pmb{\upsilon}_j$ is also influenced by the \emph{cell state} $\pmb{\zeta}_j$, both governed by
\begin{align*}
    \pmb{\zeta}_j &= \mathbf{f}_j \odot \pmb{\zeta}_{j-1} + \mathbf{i}_j \odot \mathbf{c}_j \\
    \pmb{\upsilon}_j &= \mathbf{o}_j \odot \tanh(\pmb{\zeta}_j)
\end{align*}
where the $\odot$ operator denotes the Hadamard product (element-wise product).
Vectors $\pmb{\zeta}_{j-1}$ and $\pmb{\upsilon}_{j-1}$ jointly determine the context information for the subsequent cell $j$, influencing its output.
Therefore, in the defined scheduling, only the initial values $\pmb{\zeta}_0$ and $\pmb{\upsilon}_0$ of the first AP are set to zero vectors, while all others depend on the output of the preceding subchain.

Although the entire process can certainly also occur in a centralized manner, in a fully decentralized scheme, this approach also helps minimize fronthaul signals; as each AP sends only one signal (containing context information) exclusively to the next AP in the sequence, potentially bypassing the CPU altogether.
In other words, this approach ensures that each AP receives context information from the preceding one, processes the relevant user cells, and then forwards the updated context to the subsequent AP in the sequence, as evident from Fig.~\ref{fig:DL}.

The output vector from each cell is subsequently fed into a multi-layer \emph{Fully Connected} (FC) network with an output layer size of $L$, over which sigmoid activation functions are applied.
This network is shared across inputs, forming a unified block with the underlying cell that is identical for each AP.

Considering the output size, the entire network effectively computes the probability to activate connections between the UE and each AP.
During testing, connections are determined directly by this probability, activating any for which it exceeds $50\%$.
The training phase, however, is more complex, because it uses the probabilities to determine an equal number of Bernoulli distributions.
Based on them, activations are randomly generated for calculating both their number and the corresponding SE.
The gradient of the loss function \eqref{eq:objectiveProblem} with respect to these random inputs is then also backpropagated through the network to update the weights.
Note that, during training, the LSTM weights are updated based on gradients coming from both the upward direction and subsequent cells.

\begin{figure}[!t]
    \centering
    \includegraphics[width=0.5\linewidth]{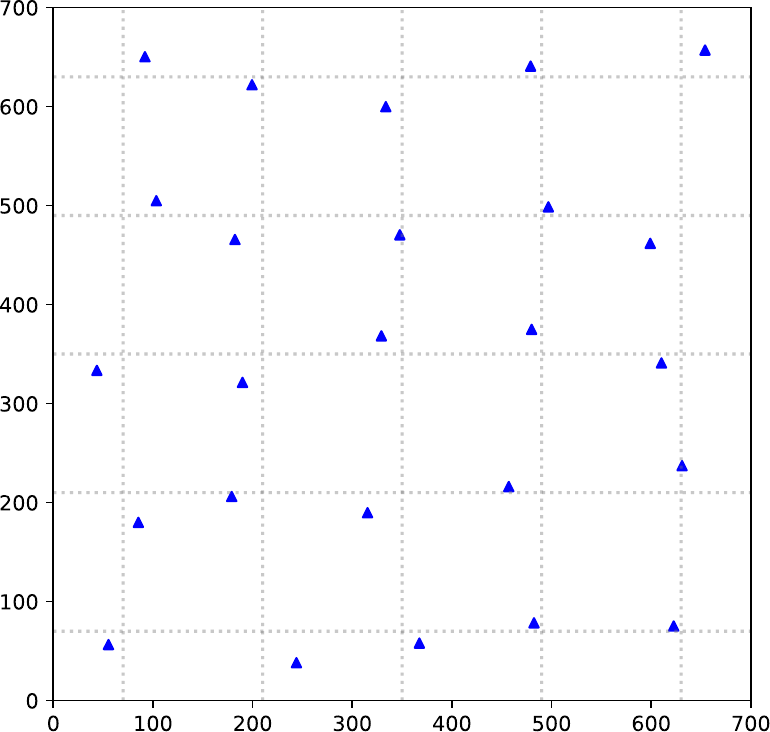}
    \caption{Example of AP positions within the area, with random variability on both axes compared to the points of an underlying imaginary grid.}
    \label{fig:area}
\end{figure}

Training is also performed in mini-batches, where the batch size determines the number of large-scale fading configurations generated based on fixed positions. 
In other words, for each location of the single UE in each training epoch, a number of fading configurations equal to the batch size are randomly produced.

Finally, the two constraints \eqref{eq:C1Problem} and \eqref{eq:C2Problem} are met by ensuring that each UE is always connected to its master AP and that, once the cluster has been defined, the power associated with each UE connected to AP $\ell$ is defined as in \cite{Interdonato2019}, i.e. by setting
\begin{equation}
    \rho_{k\ell} = \rho_\mathrm{max} \frac{\sqrt{\beta_{k\ell}}}{\displaystyle\sum_{i \in \mathcal{K}_\ell} \sqrt{\beta_{i\ell}}}
\end{equation}
where $\rho_\mathrm{max}$ is the maximum power available to the AP, which is assumed to be the same for all APs.

\section{Simulation results}
We consider a setup with $L=25$ APs, each equipped with $N=4$ antennas, arranged in a $700 \times 700$ m\textsuperscript{2} area. 
The locations are created by randomly displacing positions relative to an imaginary grid, allowing random variations up to $50\%$ of the distance between the APs, as illustrated in Fig.~\ref{fig:area}.
In the same area, multiple uniformly distributed locations are generated for the $K=10$ single-antenna UEs to form the training and test sets.
In particular, for each UE, we generate $1000$ locations for the training set and $200$ for the test (never used in training), from which we derive the subsequent results.

Regarding the large-scale fading coefficient between AP $\ell$ and UE $k$, we have 
$\beta_{k\ell}^{[\mathrm{dB}]} = \mathrm{SF}^{[\mathrm{dB}]}_{kl} - \mathrm{PL}^{[\mathrm{dB}]}_{kl}$, where $\mathrm{SF}^{[\mathrm{dB}]}_{kl}$ is the shadow fading and $\mathrm{PL}^{[\mathrm{dB}]}_{kl}$ represents the path loss, in logarithmic units.
For NLoS scenarios in urban environments, the path loss was determined based on the 3GPP microcell model \cite[Table B.1.2.1-1]{36.814} (valid from 2 to 6 GHz) 
\begin{equation}
    \mathrm{PL}^{[\mathrm{dB}]}_{k\ell} = 36.7 \log_{10} d_{k\ell} + 22.7 + 26 \log_{10} f_c
\end{equation}
where $d_{k\ell}$ denotes the distance from UE $k$ to AP $\ell$ (in meters), and $f_c$ represents the carrier frequency (in GHz).
In this context, shadow fading $\mathrm{SF}_{k\ell}$ is assumed to follow a lognormal distribution, such that $\mathrm{SF}^{[\mathrm{dB}]}_{k\ell} \sim \mathcal{N}(0, \sigma_{\mathrm{SF}}^2)$, with the standard deviation set to $\sigma_{\mathrm{SF}} = 4$ dB \cite{36.814}.

Although uncorrelated shadowing between different APs is expected, since they refer to large distances and/or different directions, a significant correlation may exist between the terms from different UEs to the same AP. 
This is taken into account through the following exponential function
\begin{equation}
	\mathbb{E}\{\mathrm{SF}^{[\mathrm{dB}]}_{i\ell} \mathrm{SF}^{[\mathrm{dB}]}_{j\ell}\} = \sigma_{\mathrm{SF}}^2 2^\frac{-d_{ij}}{\delta_\mathrm{SF}}
\end{equation}
where $d_{ij}$ is the distance between the UEs (in meters), and $\delta_\mathrm{SF}$ is the so-called ``correlation length'', which depending on the environment is hereinafter set to $9$ m \cite[Table B.1.2.2.1-4]{36.814}.

The network used has a hidden LSTM state size of $512$ and a shared FC network consists of three layers, with $256$, $128$, and $25$ neurons respectively.
In our simulations, as typically done, we set the pre-log factor without considering the data uplink phase (i.e., $\tau_u = 0$).
All other simulation parameters are listed in Table~\ref{tab:param}, where it is noted that the noise is assumed to be consistent in both the uplink and downlink (i.e., $\sigma_\mathrm{ul}^2 = \sigma_\mathrm{dl}^2$).

\begin{table}[!t]
    \caption{Simulation parameters}
    \centering
    \begin{tabular}{r|l}
        \textbf{Parameter} & \textbf{Value} \\ \hline
        Bandwidth & $20$ MHz \\
        Carrier frequency & $2$ GHz \\
        Samples per block $\tau_c$ & $200$ \\
        Noise power $\sigma_\mathrm{ul}^2 = \sigma_\mathrm{dl}^2$ & $-94$ dBm \\
        Per-UE uplink power $\eta_k$ & $100$ mW \\
        Per-AP maximum downlink power $\rho_\mathrm{max}$ & $200$ mW \\
        Difference in height between AP and UE & $10$ m \\
        Number of training epochs & $200$ \\
        Lagrange multiplier $\lambda$ & $0.04$ \\
        Learning rate & $0.00001$\\
        Batch size & $64$
    \end{tabular}
    \label{tab:param}
\end{table}

The simulations conducted were compared with the approach presented in \cite{Bjornson2020}, which mitigates interference between pilots by selecting the best channel for each available pilot across all APs.
We performed two sets of simulations to assess the efficacy of the proposed method, considering scenarios both with and without pilot interference.
The first scenario introduces strong interference with $\tau_p = 3$, whereas the second scenario virtually eliminates this interference by setting $\tau_p=10$, matching the number of UEs.

\begin{table}[!b]
    \caption{Average results.}
    \centering
    \begin{tabular}{r|c|c|c|c}
        \textbf{Orthogonal} & \multicolumn{2}{|c|}{\textbf{Strategy in} \cite{Bjornson2020}} & \multicolumn{2}{|c}{\textbf{Our approach}}\\ 
        \cline{2-5} 
        \textbf{pilots} & \textbf{\textit{SE sum}} & \textbf{\textit{Connections}} & \textbf{\textit{SE sum}} & \textbf{\textit{Connections}}\\ \hline
        $\tau_p = 3$ & $\pmb{24.42}$ & $75$ & $23.94$ & $\pmb{32.60}$ \\
        $\tau_p = 10$ & $24.65$ & $250$ & $\pmb{25.42}$ & $\pmb{38.18}$
    \end{tabular}
    \label{tab:perfor}
\end{table}

The average results on the test set are presented in Table~\ref{tab:perfor}, where as usual the sum of the SE is in bit/s/Hz.
From this table, it's evident that the network markedly reduces the average connection count, obtaining average total SE values that not only closely align with those of the classical approach but also exceed it in the absence of shared pilots.
Note that interference due to pilot sharing is challenging for the network to eliminate, as it operates without direct knowledge of the assigned pilot.

\begin{figure}[!t]
    \centering
    \subfloat[][]{\includegraphics[width=.55\linewidth]{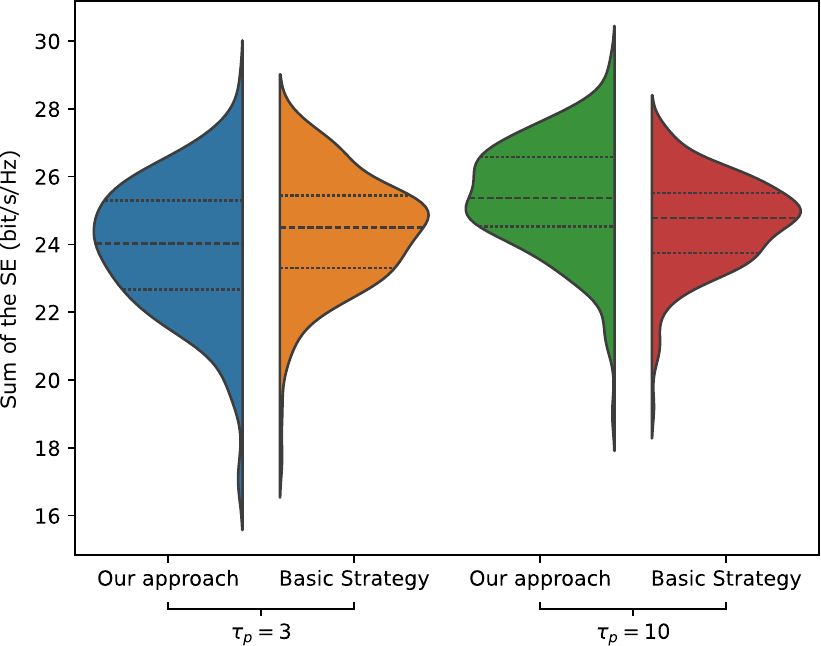}\label{fig:SE}}
    \hfill
    \subfloat[][]{\includegraphics[width=.33\linewidth]{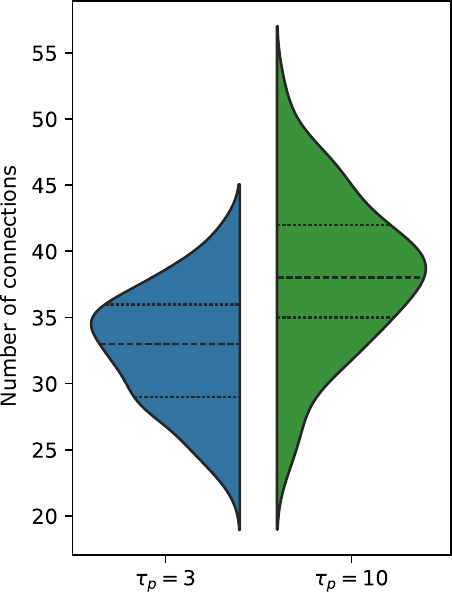}\label{fig:Conn}}
    \caption{Statistics for both scenarios relating: (a) the distribution of the sum of SE values compared to the approach outlined in \cite{Bjornson2020}, and (b) the distribution of the number of connections activated by our approach.}
    \label{fig:stats}
\end{figure}

The statistical data are also presented in greater detail in Fig.~\ref{fig:stats}.
In particular, Fig.~\ref{fig:SE} show how the network sometimes outperforms the classical approach even under high interference, while Fig.~\ref{fig:Conn} illustrates the network's adaptability to pilot interference by increasing the number of connections when it is possible to achieve higher SEs.

\begin{figure}[!b]
    \centering
    \subfloat[][]{\includegraphics[width=.45\linewidth]{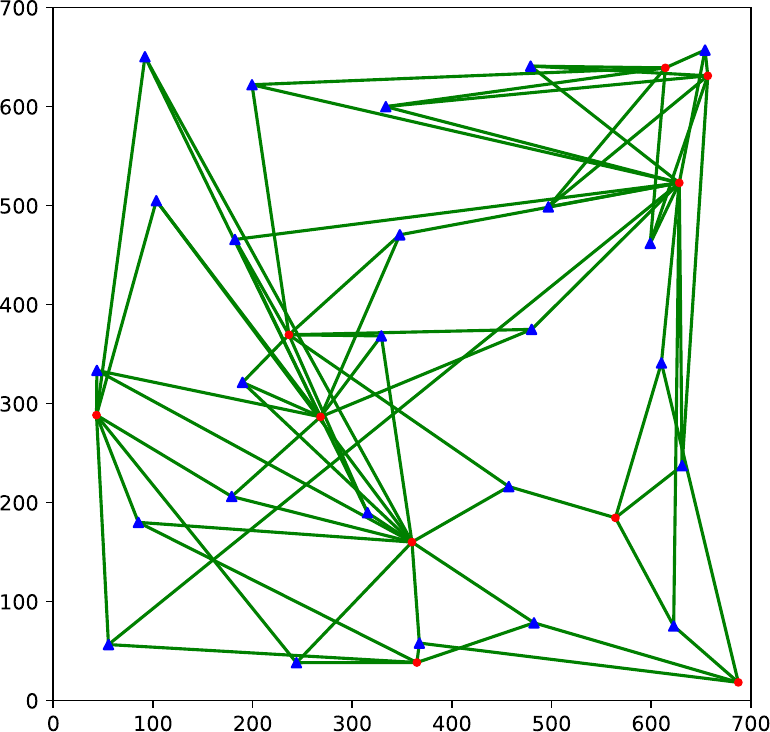}\label{fig:E1S}}
    \hfill
    \subfloat[][]{\includegraphics[width=.45\linewidth]{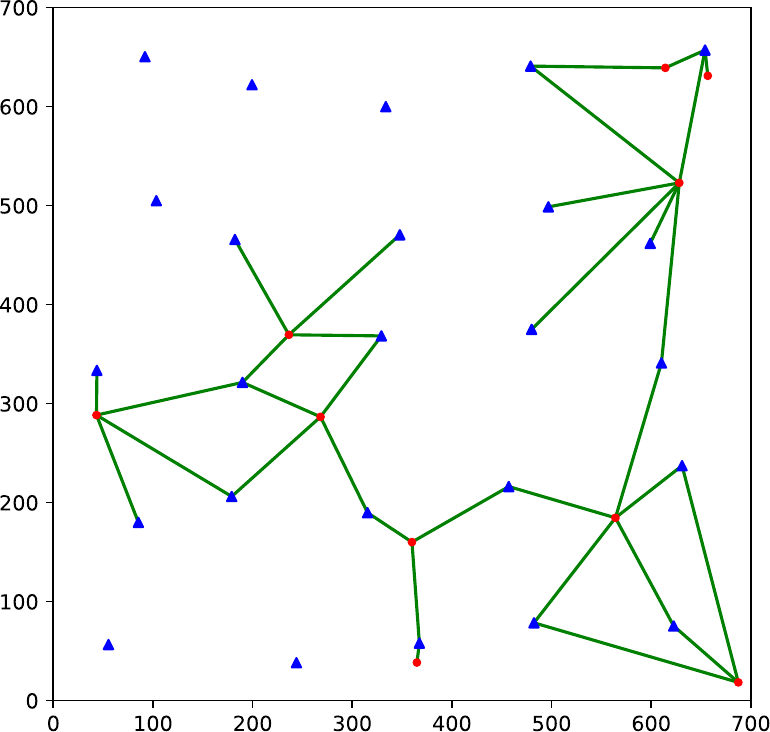}\label{fig:E1N}}
    \caption{Example illustrating connections between UEs and APs with $\tau_p=3$: (a) as per the approach outlined in \cite{Bjornson2020}, and (b) our approach after training.}
    \label{fig:example1}
\end{figure}

\begin{figure}[!t]
    \centering
    \subfloat[][]{\includegraphics[width=.45\linewidth]{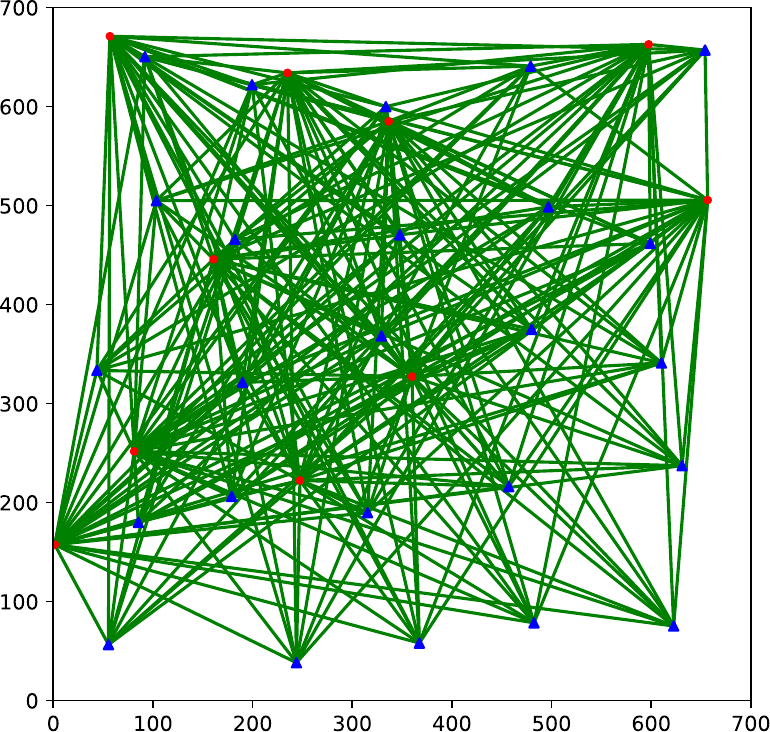}\label{fig:E2S}}
    \hfill
    \subfloat[][]{\includegraphics[width=.45\linewidth]{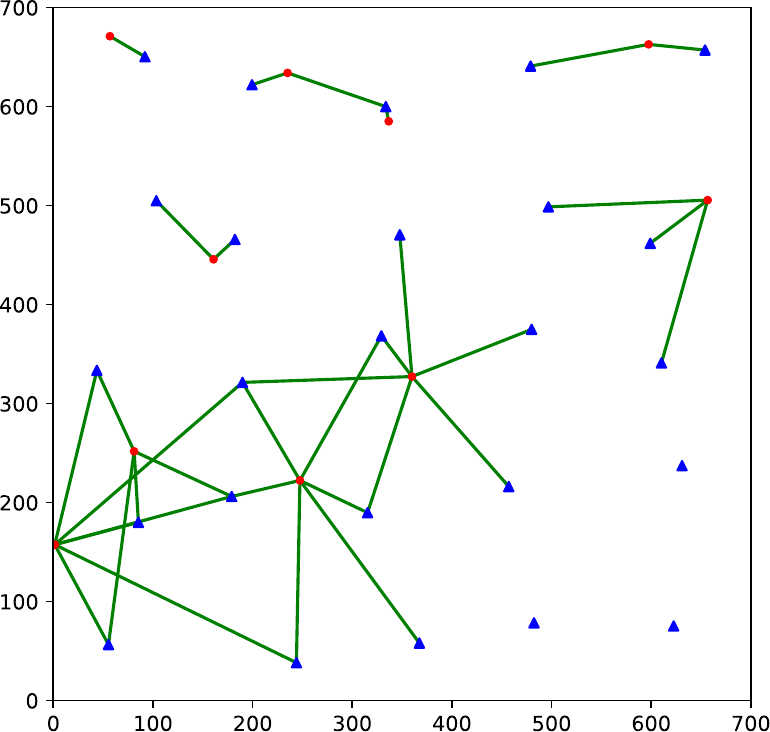}\label{fig:E2N}}
    \caption{Example illustrating connections between UEs and APs with $\tau_p=10$: (a) as per the approach outlined in \cite{Bjornson2020}, and (b) our approach after training.}
    \label{fig:example2}
\end{figure}

To further highlight the potential of the adopted approach, we selected two locations from the test set that exhibit performances closely aligned with the averages of the sum of SE for the two reference scenarios, highlighting the active connections. 
As depicted in Fig.~\ref{fig:E1N}, with $\tau_p=3$ our proposed approach establishes only $33$ connections, effectively deactivating many APs and thus significantly reducing power consumption. 
In contrast, the classical approach of Fig.~\ref{fig:E1S} activates all APs, establishing $75$ links by connecting to three optimal channels with as many UEs.
Nevertheless, the difference in the total SE is minimal: the classical approach achieves $23.84$ bit/s/Hz compared to our $23.42$ bit/s/Hz.

For $\tau_p=10$, as depicted in Fig.~\ref{fig:example2}, similar considerations apply. 
However, a notable difference emerges considering that, in this case, the classical approach connects all APs and UEs (resulting in $250$ links). 
Despite this, due to the use of MR precoding, this approach yields a lower total SE of $24.63$ bit/s/Hz, while our approach achieves a higher sum of SE, equal to $25.39$ bit/s/Hz, still using only $33$ connections.

\section{Conclusions}
The problem of AP-UE association in CF-mMIMO networks has been addressed in this paper.
A scalable deep learning approach based on LSTM networks has been introduced in order to  determine the AP-UE association maximizing the sum spectral efficiency while minimizing the number of active connections. Numerical results have shown the superiority of the proposed approach with respect to classical heuristics-based competing alternatives. 
Further research in this area is devoted to more extensive numerical analysis of the suggested strategy, extending it to scenarios optimizing different objective functions (e.g., energy efficiency) and broadening the optimization to also include transmitted power.

\bibliographystyle{ieeetr}
\bibliography{IEEEabrv, biblio}

\end{document}